\documentclass[aps,prl,nofootinbib,twocolumn,showpacs,superscriptaddress,reprint]{revtex4-1}

\usepackage{graphicx}
\usepackage{epsfig}
\usepackage{amsmath}
\usepackage{xcolor}
\usepackage[normalem]{ulem}  

\newcommand{\pdct}[3]{\left.{\partial #1}/{\partial #2}\right|_{#3}}
\newcommand{\pdc}[3]{\left.{\frac{\partial #1}{\partial #2}}\right|_{#3}}

\begin{document}

\title{Hot third family of compact stars and the possibility of core-collapse supernova explosions}

\author{Matthias Hempel}
\email{matthias.hempel@unibas.ch}
\author{Oliver Heinimann}
\affiliation{University of Basel, 4058 Basel, Switzerland}
\author{Andrey Yudin} 
\affiliation{Institute of Theoretical and Experimental Physics, 117218 Moscow, Russia}
\author{Igor Iosilevskiy}
\affiliation{Joint Institute for High Temperatures (Russian Academy of Sciences), 125412 Moscow, Russia}
\author{Matthias Liebend\"orfer}
\author{Friedrich-Karl Thielemann}
\affiliation{University of Basel, 4058 Basel, Switzerland}

\date{\today}

\begin{abstract}
A phase transition to quark matter can lead to interesting phenomenological consequences in core-collapse supernovae, e.g., triggering an explosion in spherically symmetric models. However, until now this explosion mechanism was only shown to be working for equations of state that are in contradiction with recent pulsar mass measurements. Here, we identify that this explosion mechanism is related to the existence of a third family of compact stars. 
For the equations of state investigated, the third family is only pronounced in the hot, early stages of the protocompact star and absent or negligibly small at zero temperature and thus represents a novel kind of third family. This interesting behavior is a result of unusual thermal properties induced by the phase transition, e.g., characterized by a decrease of temperature with increasing density for isentropes, and can be related to a negative slope of the phase transition line in the temperature-pressure phase diagram.
\end{abstract}

\maketitle
\section{Introduction}
The explosion mechanism of core-collapse supernovae (CCSNe) is a long-standing problem in astrophysics.
The occurrence of quark matter in CCSNe can have interesting consequences in this respect, see, e.g., \citet{takahara85,gentile93,drago99,yasutake07}.
Some other scenarios involve absolutely stable strange quark matter (see, e.g., Refs.~\cite{benvenuto89,ouyed02,ouyed13,drago15b}), which we do not consider in the present study.
\citet{sagert09} simulated a CCSN in spherical symmetry with a phase transition to quark matter that sets in at rather low densities. They found that at a certain point during the accretion in the postbounce phase, when a critical fraction of quark matter is reached in the center, the nascent protocompact star loses its stability and starts to collapse. Once pure quark matter is reached in the center, the collapse halts, and a second outgoing accretion shock is formed. It is strong enough to unbind the outer layers once it merges with the standing accretion shock, resulting in an explosion. The passage of the second shock over the neutrino spheres leads to a second neutrino burst that could be measured with present-day neutrino detectors \cite{dasgupta09} and that would give an observational signature for the QCD phase transition in CCSNe. 

However, the hybrid equations of state (EOSs) applied in Ref.~\cite{sagert09} have maximum masses much below 2~M$_{\odot}$ and are thus ruled out by the recent observations of pulsar masses around 2~M$_{\odot}$ \cite{demorest2010,antoniadis2013}. In the subsequent works exploring this scenario \cite{sagert10,fischer10,fischer11,fischer12,fischer14,nakazato13,fischer14b}, explosions could not be obtained if the maximum mass was sufficiently high. 
The required stiffening of the quark EOS leads to a weaker phase transition, with less pronounced features in the CCSN simulations. With the present investigation, we are not yet able to give a definite answer to the question of whether or not the QCD phase transition is still a viable CCSN explosion mechanism. 
However, we identified that it is related to the existence of a third family of compact stars. For the equations of state investigated, the third family is of a novel kind, as it is only pronounced in the hot, early stages of the protocompact star. We show that this behavior is related to particular features of the QCD phase transition. Throughout this paper, we use units where $k_{B}=\hbar=c=1$. 

\section{Novel third family of compact stars}
Here, we consider two hybrid EOSs as representative examples. The B165 (B139) EOS uses a bag constant of $B^{1/4}=165$~MeV (139~MeV) and has a maximum mass of 1.51~M$_{\odot}$ (2.08~M$_{\odot}$). Only the former leads to explosions in spherical symmetry \cite{sagert09,fischer11,fischer14b}. The high maximum mass of the B139 EOS is achieved by the inclusion of strong interactions with coupling constant $\alpha_S=0.7$ \cite{farhi84,fischer11}. The hadronic parts of the hybrid EOSs are taken from Refs.~\cite{shen98,shen98_2} (STOS98) for B139 and from Ref.~\cite{shen11} (STOS11) for B165. STOS98 and STOS11 are based on the same underlying model, but have numerical differences. Global charge neutrality was assumed for the phase transition, and Gibbs conditions for phase equilibrium have been applied.

\label{sec:mr}
\begin{figure}
\includegraphics[width=0.8\columnwidth]{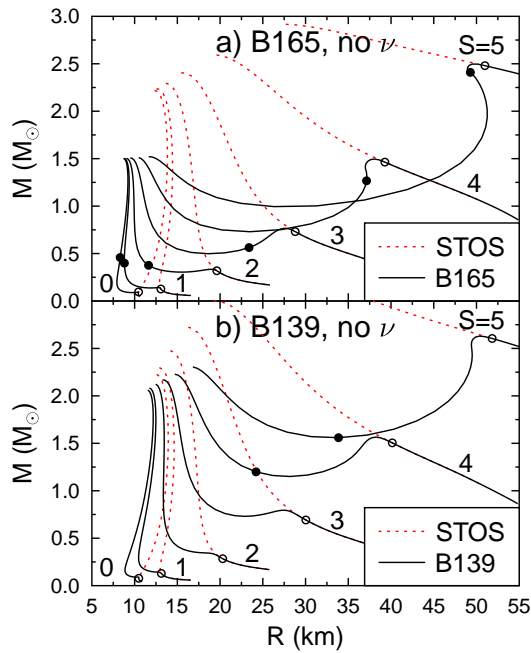}
\caption{$M$-$R$-relations for different entropies per baryon $S$ (indicated by the numbers in the figure) and in beta-equilibrium without neutrinos. a) B165 hybrid EOS (black solid lines) and STOS11 hadronic EOS (red dotted lines). b) B139 hybrid EOS (black solid lines) and STOS98 hadronic EOS (red dotted lines). Open (full) circles indicate that phase coexistence (pure quark matter) has been reached.} 
\label{fig:mr_beta}
\end{figure}
\begin{figure}
\includegraphics[width=0.8\columnwidth]{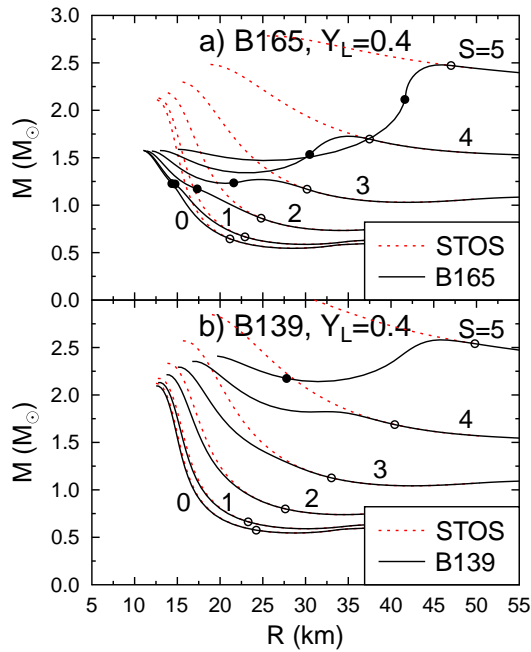}
\caption{As Fig.~\ref{fig:mr_beta}, but with completely trapped neutrinos in beta equilibrium and a lepton fraction $Y_L$ of $0.4$.}
\label{fig:mr_yl0.4}
\end{figure}
In Fig.~\ref{fig:mr_beta}, we show the mass-radius relations of the four EOSs, for various entropies per baryon $S$ and in beta equilibrium without neutrinos.\footnote{We terminate the numerical integration of the Tolman-Oppenheimer-Volkoff equations at a pressure of $10^{30}$~erg/cm$^3$, to avoid an unrealistically large contribution of a hot, low-density envelope.} 
One observes an interesting feature for the hybrid EOSs: with increasing entropy, a second maximum develops, which eventually even becomes the global maximum. It is well known from cold compact stars that stars on the branch between the right maximum and the local minimum are unstable with respect to radial perturbations \cite{schertler00}. In such a situation, where one has two different stable branches in addition to the one of white dwarfs, one speaks about a third family of compact stars \cite{gerlach68,schertler00,Glen00}. Solutions with different radii at equal masses are also called twins \cite{schertler00,Glen00}. In Fig.~\ref{fig:mr_beta}, we find the third family feature to depend on entropy; similarly as in Ref.~\cite{yudin13}. It does not exist or is only very tiny for cold compact stars and becomes extremely pronounced for increasing entropy. We are not aware that this possibility has been identified in the literature before.

For B165, the third family branch appears for lower entropies and is generally more pronounced than for B139. 
It is also interesting that B139 hybrid stars do not contain pure quark matter, except at highest entropies.
Note that for both EOSs for $S=5$ there is a third family, but its maximum total number of baryons is below the one of the second family. In this case, the third family cannot be reached by accretion, and a collapse from the maximum of the second family would end in a black hole as the number of baryons is conserved during a collapse.

In protocompact stars, one not only has a finite entropy but also a finite fraction of neutrinos, which are typically completely trapped in the core. To identify their effect on the stability of protocompact stars, in Fig.~\ref{fig:mr_yl0.4}, we consider completely trapped neutrinos and a constant lepton fraction $Y_L$ of $0.4$. Neutrinos tend to decrease effects of the phase transition and the third-family feature, because they give a similar contribution to the thermodynamic properties in both phases. In Fig.~\ref{fig:mr_yl0.4} (a), the third family is only visible for $S\geq3$, and in Fig.~\ref{fig:mr_yl0.4} (b), it is only visible for $S\geq4$. However, it is not very realistic to assume a constant value of $Y_L$ throughout the protocompact star. The results shown in Fig.~\ref{fig:mr_yl0.4} (Fig.~\ref{fig:mr_beta}) can be considered as an overestimate (underestimate) of the effect of neutrinos. A more realistic situation should be somewhere in between. 

\section{Relation to the collapse}
\citet{fischer11} gave detailed explanations of the processes before and during the collapse of the protocompact star in the CCSN. Here, we complement them by connecting this collapse with the existence of a third family of protocompact stars as shown in Figs.~\ref{fig:mr_beta} and \ref{fig:mr_yl0.4}. 
In CCSNe, in the absence of shocks and as long as neutrinos are trapped, $S$ and $Y_L$ are conserved quantities that are only advected with the matter, and their values are similar as in Figs.~\ref{fig:mr_beta} and \ref{fig:mr_yl0.4}. 
During the ongoing accretion in the postbounce phase, the mass and central density of the protocompact star increase continuously.  
One moves along an $M$-$R$-curve to the left. Either the phase transition onset has appeared already around bounce or it is reached during this accretion phase. Initially, it does not have a significant effect on the dynamics \cite{fischer11}. This changes when
the central density has increased so far that the mostly hadronic protocompact star reaches the maximum mass of the second family. At this point, a collapse is induced. It proceeds until high enough pressures on the third family branch that counterbalance gravity are reached. Afterward, the second shock is formed that has the potential to explode the star; see Ref.~\cite{fischer11}. 

Even though the kinetic energy of the protocompact star collapse is not directly converted into explosion energy, see Ref.~\cite{fischer11}, it is interesting to look at the release of gravitational binding energy\footnote{It is given by the mass difference between the maximum of the second branch and a star with equal number of baryons on the third branch.} for the third families shown in the figures above. 
It ranges from 0 to $\sim 120\times 10^{51}$~erg, where the values are increasing with $S$ and decreasing with $Y_L$. 
This range is similar or even larger as found for transitions to a third family in cold compact stars; see, e.g., Refs.~\cite{mishustin03,alvarez2015}.
For B139, the energies are generally lower than for B165, but for both EOSs, they
can exceed the typical explosion energy of a CCSN by orders of magnitude if entropies are high and the lepton fraction is low. The finding that the third family is much less pronounced for B139 is consistent with the result that this EOS did not lead to explosions in spherical CCSN simulations \cite{fischer14b}. Only a pronounced third family seems to be favorable for explosions. 

We remark that the collapse from the second to the third family of \textit{cold} compact stars has been studied already in the literature; see, e.g., Refs.~\cite{mishustin03,alvarez2015}. 
In addition to the energy release, in both cases, an accompanying neutrino and/or a gamma-ray burst is expected. \citet{pagliara09} noted that also deleptonization can trigger a collapse from the second to the third family, which represents a related scenario. A collapse in rotating stars can also lead to the emission of gravitational waves \cite{dimmelmeier09}. With our findings we confirm previous expectations that a third family can trigger CCSN explosions.

\section{Unusual thermal properties of the EOS induced by the phase transition}
In the following, we discuss which properties of the phase transition are responsible for the formation of the hot third family.
For most EOS, at fixed baryon number density $n_B$, the pressure $P$ is increasing with temperature $T$; consider, e.g., an ideal Maxwell-Boltzmann or Fermi-Dirac gas. However, it is also possible in special situations that $\pdct{P}{T}{n_B}<0$.\footnote{Note that this does not violate thermodynamic stability.} In Ref.~\cite{iosilevskiy15}, this was called ``abnormal thermodynamics,'' and it was pointed out in Refs.~\cite{iosilevskiy14,iosilevskiy15} that such an unusual sign of a second-order mixed partial derivative of the thermodynamic potential never occurs isolatedly, but is accompanied by a change of the sign of many other second-order mixed partial derivatives. For example, one has
\begin{equation}
\pdc{P}{T}{n_B}<0 \Leftrightarrow \pdc{T}{n_B}{S}<0 \; .
\end{equation}

\begin{figure}
\includegraphics[width=0.8\columnwidth]{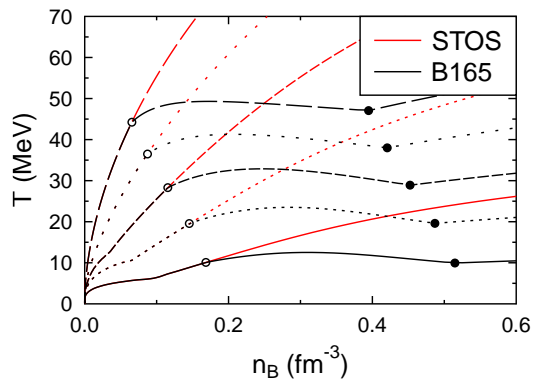}\\
\caption{Temperature as a function of baryon number density for isentropes with $S=1$, 2, 3, 4, and 5, (increasing from bottom to top) and $Y_L=0.4$ for the B165 (black solid lines) and STOS11 (red dotted lines) EOSs. The open (full) circles mark the onset (end) of the phase transition.}
\label{fig:t_nb}
\end{figure}
As is visible in Fig.~\ref{fig:t_nb}, a region with $\pdct{T}{n_B}{S}<0$ is present for all entropies. It shifts to lower densities and becomes more pronounced by increasing $S$. For the B139 EOS, such a negative slope is only found for high values of $S$ above 4. Generally, it occurs only inside the phase transition region; see Fig.~\ref{fig:t_nb}. This unusual decrease of temperature due to the phase transition has also been found for various hybrid EOSs in other works, see, e.g., Refs.~\cite{drago99,steiner00,satarov09,nakazato10,fischer11,yudin13}, but without discussing any further implications for compact stars. This special thermal property can also be present without a first-order phase transition; for example, in Ref.~\cite{masuda15}, a negative value of $\pdct{T}{n_B}{S}$ was identified for a crossover transition from hadronic to quark matter, and in Ref.~\cite{drago15a} it was identified for a hadronic EOS including hyperons and Deltas. 

Here we show that this behavior can be relevant for the stability of compact stars. The only information needed to calculate the $M$-$R$ relation of compact stars is the $P(\epsilon)$ relation. To investigate the effect of finite entropies, we therefore consider the derivative
\begin{equation}
\pdc{P}{S}{\epsilon}=-Tn_Bc_s^2 +\frac{T}{C_V}\pdc{P}{T}{n_B} \; , \label{eq_dpds}
\end{equation}
where $\epsilon$ is the energy density, $c_s$ is the speed of sound, and $C_V$ is the heat capacity per baryon (see also Ref.~\cite{yudin15}). 
If $\pdct{P}{S}{\epsilon}$ is positive (negative), it corresponds to a stiffening (softening) of the EOS with increasing $S$. The first term, which is a relativistic correction, is always negative. 
For both EOSs of the single phases, the second term is always positive and larger than the first one, and thus one has stiffening. However, for matter in the phase transition region with $\pdct{T}{n_B}{S}<0$, or equivalently $\pdct{P}{T}{n_B}<0$, and because $C_V>0$, $\pdct{P}{S}{\epsilon}$ will be negative, and one has softening.

\begin{figure}
\includegraphics[width=0.8\columnwidth]{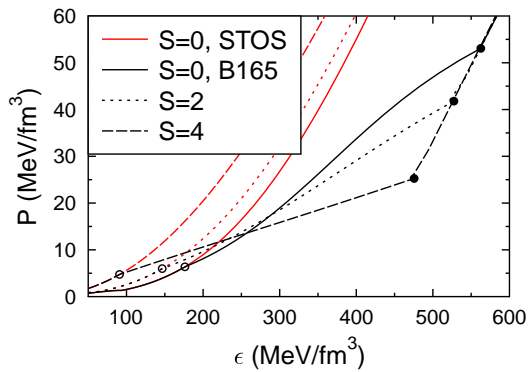}
\caption{The pressure as a function of energy density for the B165 (black lines) and STOS11 (red lines) EOSs for $Y_L=0.4$ and various entropies per baryon $S$. The open (full) circles mark the onset (end) of the phase transition.} 
\label{fig:pe_b165}
\end{figure}
This is illustrated in Fig.~\ref{fig:pe_b165}, where one sees that 
in the hadronic phase the pressure is always increasing with entropy. In the pure quark phase, the pressure is only slightly increased by entropy. Outside the phase transition, one thus has the normal behavior that the EOS is stiffened when it is heated. 
However, in a part of the phase transition around $\epsilon = 400$~MeV/fm$^{3}$, one has a significant softening (see also Ref.~\cite{satarov09}); i.e., the pressure is decreasing with entropy, which is due to a large negative value of $\pdct{P}{T}{n_B}$. 

The stiffening in the single phases and softening in the phase transition region is responsible for the behavior of the $M$-$R$ relations discussed above: on the one hand, overall, the maxima of the second and, if present, also of the third family are increasing with entropy. On the other hand, stars of which the central part has just entered the phase transition get unstable if entropies are sufficiently high. For the two EOSs discussed here, one can conclude that the unusual thermal properties of the phase transition, characterized, e.g., by  $\pdct{T}{n_B}{S}<0$, are responsible for the observed special third-family features.

\section{Relation to the QCD phase diagram}
Let us consider symmetric nuclear matter without strangeness. In this case, the pressure in the coexistence region of the phase transition is solely a function of temperature,\footnote{In Ref.~\cite{hempel13}, this was denoted as a ``congruent phase transition,'' using the terminology of Ref.~\cite{iosilevskiy10}.} and thus we can identify $\pdct{P}{T}{n_B}=\left.{dP}/{dT}\right|_{\rm PT}$, where the latter quantity denotes the slope of the phase transition line in the temperature-pressure phase diagram. By using Eq.~(\ref{eq_dpds}), one can thus relate the QCD phase diagram with a possible softening or stiffening of the EOS with increasing entropy. 

In Refs.~\cite{iosilevskiy14,hempel13,iosilevskiy15,steinheimer14}, it was shown that the slope $\left.{dP}/{dT}\right|_{\rm PT}$ is negative for the QCD phase transition and positive for the liquid-gas phase transition of nuclear matter (see also Refs.~\cite{satarov09,bombaci09,nakazato10}). This qualitative difference can be used as a subclassification of first-order phase transitions; in Refs.~\cite{iosilevskiy14,iosilevskiy15}, they were called \textit{entropic} and \textit{enthalpic}, respectively. 
For possible experimental signatures of this property in heavy-ion collisions, see Ref.~\cite{steinheimer14}.
Also, for the hybrid EOSs employed in the present study, we have found a negative slope, i.e., that the QCD phase transition is entropic. For such a phase transition one always has the unusual thermal properties outlined above (e.g., $\pdct{T}{n_B}{S}<0$) and a significant softening of the EOS with increasing entropy if $\left.{dP}/{dT}\right|_{\rm PT}$ is sufficiently negative. We remark that the general relation between the slope of the phase transition line and the unusual behavior of the second-order mixed partial derivatives of the thermodynamic potential was first noted in Ref.~\cite{iosilevskiy14}.

The reason for this special property of the QCD phase transition can be identified by using the Clausius-Clapeyron equation \cite{bombaci09,iosilevskiy14,yudin13,hempel13,iosilevskiy15},
\begin{equation}
 \left.\frac{dP}{dT}\right|_{\rm PT}=\frac{S^I-S^{II}}{1/n_B^I-1/n_B^{II}} \; ,
\label{clapeyron}
\end{equation}
where $I$ and $II$ denote the two phases in coexistence and we assume $n_B^I < n_B^{II}$ in the following. One has $S^I<S^{II} \leftrightarrow \left.{dP}/{dT}\right|_{\rm PT}<0$. The QCD phase transition has a negative slope (i.e., $\left.{dP}/{dT}\right|_{\rm PT}<0$) because the quark phase has a higher entropy per baryon and specific heat than the hadronic phase, which can also be inferred from Fig.~\ref{fig:t_nb}. In the liquid-gas phase transition the opposite is the case: the denser phase (the liquid) has the lower entropy per baryon, $S^{II}<S^I$. 

We remark that the hybrid EOSs considered in the present study contain strange quarks in weak equilibrium and a leptonic component to maintain charge neutrality, and generally we consider isospin asymmetric systems. As a consequence, it is not possible to relate $\pdct{P}{S}{\epsilon}$ directly with the slope of a phase transition line and only parts of the coexistence region of the B165 and B139 EOSs show the unusual thermal properties. 

\section{Summary and conclusions}
In the present study, we have investigated effects of the QCD phase transition in CCSNe. We found that the explosions reported in Refs.~\cite{sagert09,fischer11} can be explained as a transition from a second to a third family of compact stars. From this, we arrive at the general conclusion that a third family of compact stars can trigger CCSN explosions in realistic, spherically symmetric CCSN simulations, confirming previous expectations.
Interestingly, for the two hybrid EOSs considered here, the third-family feature was very tiny in the case of cold compact stars but extremely enhanced with increasing entropy. We are not aware that such a behavior, where the third family is only pronounced at the early, hot protocompact star stage, has been identified in the literature before.

This novel third family was explained to be a result of unusual thermal properties of the EOS induced by the phase transition. The unusual thermal properties are characterized by a decrease of temperature with density along isentropes, $\pdct{T}{n_B}{S}<0$; a negative slope of the phase transition line in the temperature-pressure plane, $\left.{dP}/{dT}\right|_{\rm PT}<0$; and higher entropies per baryon in the quark than in the hadronic phase. Such a phase transition is called ``entropic''. These properties imply a softening of the EOS with increasing entropy, $\pdct{P}{S}{\epsilon}<0$, so that the unusual thermal properties of the phase transition also favor unusual behavior in the $M$-$R$ relation. 

A negative value of $\left.{dP}/{dT}\right|_{\rm PT}$ does not automatically imply that a CCSN explosion can be triggered or that a third family can be formed. Nevertheless, from our perspective, it is quite remarkable that, at least for the two hybrid EOSs considered here, the $M$-$R$ relation and CCSN explosions can be linked with the phase diagram of QCD, the structure of which is one of the key issues in the physics of strongly interacting matter. It would be very interesting if the slope of the phase transition line could be constrained by heavy-ion collisions or lattice QCD calculations in the future; see, e.g., Ref.~\cite{steinheimer14}. If a second neutrino burst from a CCSN (with the particular properties discussed in Refs.~\cite{sagert09,fischer11}) was observed in the future, this would be a signature for the existence of a third family and the presence of quark matter in compact stars and would also be an indication of the QCD phase transition being entropic.

We close with a few comments regarding the question of whether one can have a sufficiently strong third-family feature in hot protocompact stars to trigger explosions as in Ref.~\cite{sagert09} and a sufficiently high maximum mass of cold compact stars at the same time. Of course, it is favorable if there is already a third family for cold compact stars, as discussed, e.g., in Refs.~\cite{alford2013,benic15,heinimann16}.
An extended phase-coexistence region, as seen for the B139 EOS, tends to reduce effects of the phase transition. This could be modified by the assumption of local charge neutrality. Besides that, one should explore different models for hadronic and quark matter and focus in particular on their thermal properties, which is an important insight from the present study. We have also identified that the entropy and lepton fraction profile is crucial for this explosion mechanism. Thus, it would be very interesting to investigate various different progenitors to see if some of them are particularly favorable.

Even if explosions due to the phase transition cannot be obtained in spherical symmetry, this could be possible in multidimensional hydrodynamic simulations of CCSNe (see Refs.~\cite{janka12,burrows13} and references therein), which has not been investigated so far. There might also be other interesting effects of the QCD phase transition; for example, \citet{yudin15} recently reported that the unusual thermal properties of the QCD phase transition can induce a special convective instability.

\begin{acknowledgments}
The authors would like to thank J.~Schaffner-Bielich, G.~Pagliara, I.~Sagert, and T.~Fischer for their useful comments and discussions, and I.~Sagert for providing the hybrid EOS tables used in this work. The research leading to this paper has received funding from the Swiss National Science Foundation and the European Research Council (ERC) under the European Union's Seventh Framework Programme (FP7/2007-2013) / ERC Advanced Grant Agreement N° 321263 - FISH. Partial support comes from ``NewCompStar,'' COST Action MP1304. I.I.\ acknowledges support of Russian Scientific Fund, Grant No.\ 14-50-00124.
\end{acknowledgments}

\bibliographystyle{apsrev}
\bibliography{literat}

\appendix

\end{document}